\documentclass[aps,prb,preprint]{revtex4}
\usepackage{graphicx}
\usepackage{color}
\newcommand\beq{\begin{equation}}
\newcommand\eeq{\end{equation}}
\newcommand\bear{\begin{eqnarray}}
\newcommand\eear{\end{eqnarray}}
\begin{document}

\title {Negative Differential Conductance in Nano-junctions: A
  Current Constrained Approach}

\author{Prakash Parida,$^1$ Swapan K Pati,$^{1,2}$ and Anna Painelli$^3$}

\affiliation{ $^1$Theoretical Sciences Unit, $^2$New Chemistry Unit, Jawaharlal Nehru Centre For
Advanced Scientific Research, Jakkur Campus, Bangalore-560064, India\\
$^3$Dipartimento Chimica GIAF, Parma University, \& INSTM UdR-Parma, 43100 Parma, Italy}


\begin{abstract}
\parbox{6in}
{A current constrained approach is proposed to   calculate
negative differential
conductance in molecular nano-junctions. 
A four-site junction is considered where  a steady-state
current is forced by inserting only the  two central sites within the
circuit. The two lateral sites (representing e.g. dangling molecular
groups) do not actively
participate in transport, but exchange electrons
with the two main sites. These auxiliary sites   allow for a
variable number of electrons within the junction, while, as required by
the current constrained approach, the  total number of electrons in the
system is kept constant. We discuss the conditions for  negative
differential conductance in terms of cooperativity, variability of the
number of electrons in the junction, and electron correlations.}
\end{abstract}

\maketitle


\section{Introduction} 

Understanding of charge transport through low dimensional
electronic system has got tremendous impetus in recent years
because of their huge potential in advanced electronic devices. \cite{Ratner,Sanvito}
Especially, the possibility to switch  a molecule between off and on states and to
support  decreasing current with increasing voltage, the so called negative
differential conductance (NDC) behavior, are  hot topics in
the field of single molecule electronics, mainly due to its  
credible applications. Since its first observation in the tunnel
diode by Esaki in 1958,~\cite{Esaki} NDC has been subject of many experimental and theoretical studies.
NDC has been observed in a variety of experimental systems\cite{Tour,Don} and has been discussed
theoretically, even if most of  theoretical 
work was based on one-electron pictures. \cite{Seminario,Cornil,rpati,lakshmi_PRB,lakshmi_jpc,Cuniberti1,ranjit1}
Recent experiments demonstrated NDC in double quantum dots junctions \cite{Ono,tarucha} and
have rekindled interest in the phenomenon, occurring in the low temperature, weak-coupling limit. 
Several theoretical studies on donor-acceptor double quantum dot
systems address strong rectification and NDC features within 
many-electron pictures. ~\cite{Thielmann, Hettler1, wegewis,Hung, Cuniberti, Aghassi,
Paulsson, Datta2,Datta3,Bhaskaran,IEEE,prakash} 
However, all available theoretical studies of NDC are  based on 
voltage constrained (VC) approaches,
where the electric current is driven through the junction 
by imposing a finite potential drop
between two applied  electrodes. Two semi-infinite electronic
reservoirs are unavoidably introduced in VC models, posing  some
fundamental problems and leading to a picture  
difficult to reconcile with correlated electrons. \cite{Kohn1,Mera,Green}

   As an alternative to VC approach, current constrained (CC) approaches have been developed to describe transport 
in both meso- and molecular junctions. CC approaches avoid any
reference to reservoirs, and quite naturally apply to correlated electrons.~\cite{Kohn1,Mera,Ng,Magnus,Kosov1,Kosov2,Burke,Anna} On  
physical grounds, a current can be forced in a circuit by driving a 
time-dependent magnetic flux through the circuit. The model, originally 
proposed by W. Kohn to describe optical conductivity in extended systems,~\cite{Kohn2} has been applied to describe 
electrical transport in mesoscopic systems,~\cite{Kohn1, Magnus} and more recently in molecular junctions.~\cite{Burke,Anna} 
Alternatively, variational techniques can be adopted, and the system
can be forced in a current-carrying non-equilibrium 
steady state introducing properly defined Lagrange-multipliers in the
Hamiltonian.  
 While offering an interesting alternative to the popular
and successful VC methods, the CC approach suffers from several
limitations that hinders its widespread use.\cite{Anna} Not explicitly accounting
for contacts, the CC approach  relies on  phenomenological models
for the relaxation dynamics to describe dissipation. In
polyatomic junctions, constraints must be explicitly enforced to avoid charge
accumulation at atomic sites within the junctions, leading to
a cumbersome numerical problem for large and asymmetric molecular
systems.\cite{Anna} Finally, in VC approaches electrons can be exchanged between
the  electrodes and the molecular junction and 
charge injection/depletion phenomena can be observed
upon changing the applied gate voltage. 
These phenomena are critical to many processes, including NDC, but
can not be described in CC approaches
where the current is forced through a closed circuit and 
the total number of electrons in the junctions is strictly constant.

Here we propose a simple and effective strategy to overcome the last
general problem inherent to CC approach. Specifically, we demonstrate that if one or more auxiliary sites
 are attached to the main current-carrying nanojunction, they can exchange electrons with the
junction, thus representing electron source or  sink.
Along these lines, we present for the first time 
a calculation of NDC within a CC-based  model. Describing NDC in an
alternative picture with respect to the widespread  VC-based models
deepens our current understanding of the
basic physics underlying this strongly non-linear phenomenon, as
required to control and optimize the performance of NDC-based devices.

\section{The model and method}
    
Previous work from two of the present authors based on the VC approach
demonstrated interesting nonlinear behavior in the  current-voltage
characteristics of two-dot systems with correlated
electron.~\cite{lakshmi_PRB,lakshmi_jpc} Taking clue from this work, here
we explore transport behavior in two-dots junctions within the CC approach. 
The study becomes interesting, especially in the context of the number of electrons within the junction, which plays 
an important role in the low-bias current-voltage characteristics. 

Fig. 1(a) shows a rough sketch of   the  model. We consider a  4 sites
Hubbard model as described by the Hamiltonian:
\begin{eqnarray}
H_0 & =&  \sum_{i=1}^4\ \epsilon_i\hat n_{i}- \sum_{\sigma=\uparrow, \downarrow} 
t_{i,i+1}(a^{\dag}_{i,\sigma}a_{i+1,\sigma}+h.c.)\nonumber\\
 &&+  \sum_{i=1}^4 U_i \hat n_{i\uparrow} \hat n_{i\downarrow}
\label{hami} 
\end{eqnarray}
\noindent
where $a_i$ ($a_i^\dagger$) annihilates (creates) an electron with
spin $\sigma$ on site $i$ and $\hat n_i=\sum_\sigma a_i^\dagger a_i$. For the sake of simplicity in the following
we set constant on-site repulsions, $U_i=U$, and    hopping integrals
$t_{ij}=t$, even if different choices have been explored. Moreover 
we consider an asymmetric junction with a weak bond: $-\epsilon_2=\epsilon_3=2$, $t=0.2$.\\

The eigen states of $H_0$ are stationary states and do not sustain any
current. To induce current through the central (2-3) sites we insert
these two sites into a closed circuit through which we impose a
current according to the CC prescription:
\begin{eqnarray}
H(\lambda)=H_0-\lambda \hat{j},
\label{hlambda}
\end{eqnarray}
\noindent
where $\hat{j}=-it_{23}\sum_{\sigma}
(a^{\dag}_{2,\sigma}a_{3,\sigma}-h.c.)$ measures the current flowing
through the bond between sites 2 and 3  (the junction region). 
Here and in the following $e$ and $\hbar$ are adopted as units for
charge and momentum, respectively.
The field $\lambda$ coupled to the current enters the  Hamiltonian as a
Lagrange multiplier, whose value is fixed by the requirement 
that a finite current $I=<G(\lambda)|\hat{j}|G(\lambda)>$ flows
through the system. 
{\color{blue} The eigenstate of $H(\lambda)$ describe electrons with a
  finite overall linear momentum which looks rather unphysical in an
  open circuit, as described by $H_0$. This paradox can be solved,
  according to Burke et al.,\cite{Burke} connecting the two ends of
  the junction (sites 2 and 3 in our model system) through a long,
  thin and ideally conducting wire so that electrons escaping from the
  right side of the junction immediately appear on the left side as to
  maintain fixed number of electrons inside the junction. 
As schematically shown in Fig. 1, the junction is then inserted in a
closed loop and this allows to assign a precise physical meaning to
$\lambda$,  that was introduced formally as a Lagrange
multiplier. Following the seminal work of Kohn,\cite{Kohn2} in fact  one
can force a current through a closed ring threading
an oscillating  magnetic field of frequency $\omega$ through the
center of the ring to generate a spatially uniform oscillating
electric field $E$ across the junction.}
At the leading order in the
field,\cite{maldague} the Hamiltonian in the presence of a
magnetic flux can be written as in Eq. \ref{hlambda} with $\lambda$
proportional to the amplitude of the vector potential.  Quite interestingly, a similar result has been recently reported  via a general and exact
projection operator technique.\cite{mukamelRMP}

The two central sites (sites 2 and 3) are inserted in a closed circuit
where the current is forced (cf Fig. 1a) 
 and hence take active part in transport.
The lateral sites (1 and 4) represent auxiliary sites: the current does not flow
through these sites, but they are connected to the circuit and,
exchanging electrons with the main sites (2 and 3), provide a
source/sink for the injection/removal of electrons within the circuit.
Being interested in NDC we set  $U >>t$: 
the model then describes four weakly connected quantum
dots. Alternatively, the same model  may represent
  a minimal model for a molecular junction with at
least one weak (poorly conjugated) bond and with two dangling groups (R and R')
connected to the current carrying skeleton as shown, as an
example in Fig. 1(b). From a chemical perspective, the groups R and
R', modeled by the auxiliary sites of Fig. 1a can  behave as electron
withdrawing/donating groups with respect to the main sites.    

If the auxiliary sites (1 and 4) are disconnected from the main circuit
($t_{12}=t_{34}=0$) the number of electrons participating in transport
is strictly constant: having inserted the two central sites in an
(ideal) closed   
circuit implies that electrons escaping from site 3 immediately
enter on 2 (or viceversa). As discussed above, the obvious constraint
of constant charge in a closed circuit hinders the description of
phenomena related to charge injection/depletion processes, that are instead
easily captured in VC approaches where electrons are exchanged between
the junction and the leads. 
However, if the auxiliary sites are  connected to the junction 
 the number of electrons within the circuit, $n= \langle \hat n_2+ \hat n_3
\rangle$, becomes a variable:   finite
$t_{12}$ and/or $t_{34}$ allow for charge injection/depletion within
the junction.
\begin{figure}
\centering
\includegraphics[scale=0.35,angle=0.0]{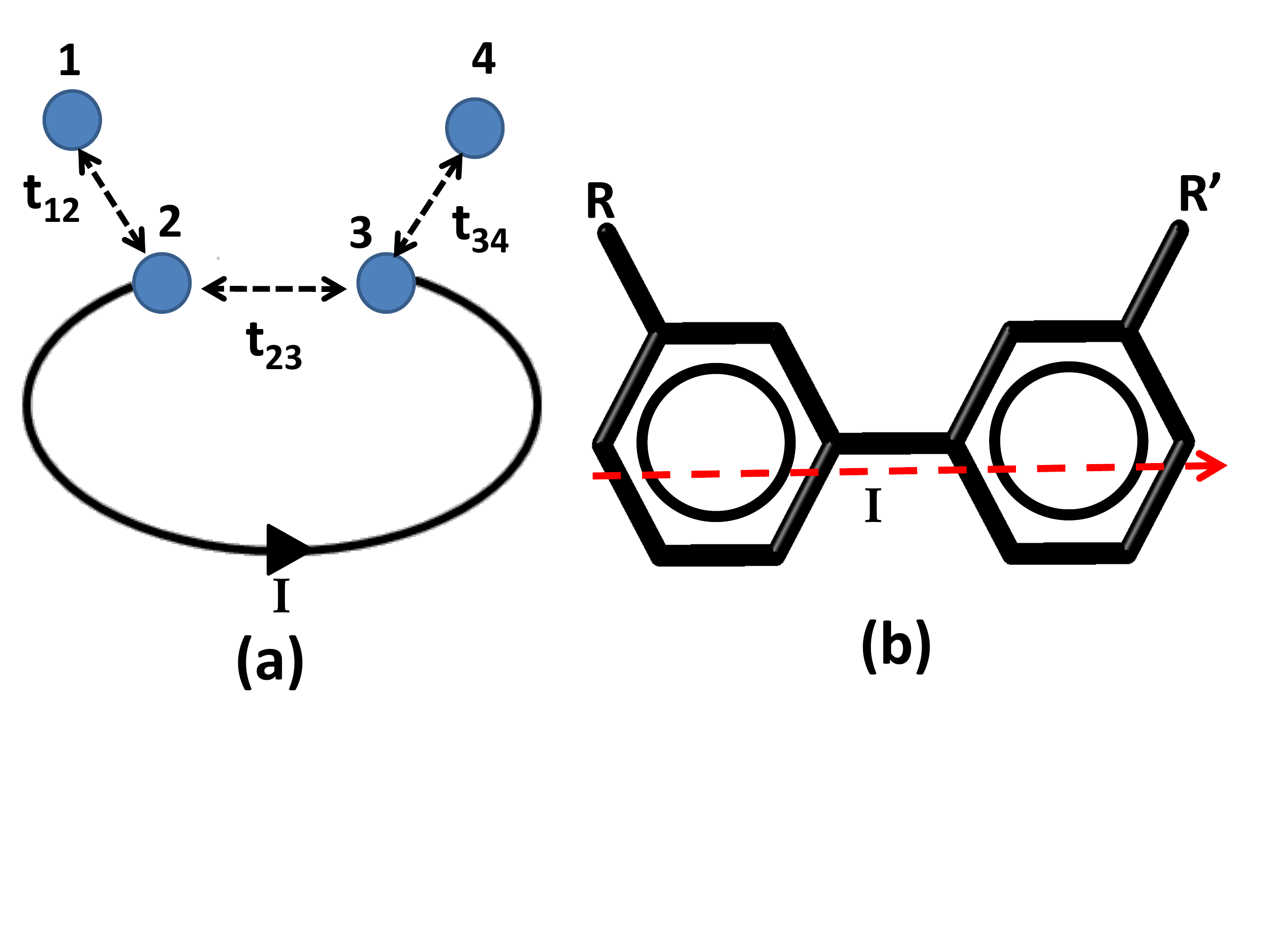}
\vspace*{-1.5cm}
\caption{\label{fig1} (a) A schematic representation of the four-site
  system, where current flows
through the two 
 central sites in the closed circuit, while the lateral sites behave as sources/sinks of electrons.
(b)The biphenyl molecule with two dangling groups, R and R', represents
an example of a  molecule that can be roughly described by the four-site
model in panel (a). The red arrows describe the direction of current
flow.}
\end{figure}

The Hubbard Hamiltonian in Eq. (1) and its current-carrying
version in Eq. (2) can be written on a real-space basis and the
resulting matrix can be diagonalized   
exactly. The real space basis  is defined by the
complete and orthonormal set of  functions ($\phi_i$) that specify the occupation of
site spin-orbitals. Both the total number of electrons, $N$, and the z-component of the 
total spin, $S_z$, commute with the Hamiltonian, and the Hamiltonian
matrix can be defined for a particular
charge and spin sector. Here we set $N=4$ and work in the subspace 
relevant to the ground state, i.e., $S_z$=0, for a grand total of 256
basis states.\\

The definition of characteristic current-voltage curves is quite subtle in CC approaches. As  discussed above, Hamiltonian in Eq. 2
describes  (at the  leading order in the vector potential)  the effect
of an oscillating magnetic flux driving a current through a
circuit. In this view, $\lambda$ is simply proportional to the
corresponding electrical field: $E\propto \omega
\lambda$.\cite{Kohn2,maldague} However in the DC limit,
$\omega\rightarrow 0$, the driving field, and hence the driving
potential vanish, leading to the unphysical result of a finite
current at zero bias. This zero-resistance picture emerges because
relaxation dynamics is neglected: in fact a finite potential drop is 
required to sustain current just to overcome  
friction in the junction. The calculation of characteristic curves in
the CC approach then requires a  model  for relaxation dynamics.             
In particular,  the current-carrying state, $|G(\lambda)>$, is a
non-equilibrium  state and the system relaxes back to the ground
state, $|G(0)>$. The
amount of power spent to sustain the system in the non-equilibrium 
current-carrying state is set by the relaxation dynamics: the faster
the relaxation the more power must be 
spent to sustain the current. The Joule law defines the
relation between the electrical power spent on the 
molecule, $W$, and the potential drop needed to sustain the current:
$W=IV$. Since $I$ is known, characteristic $I(V)$ curves 
can be obtained from $W$.\cite{Anna} 

{\color{blue} We adopt a simple phenomenological model for relaxation dynamics
that is widely used in spectroscopy. The model is general and properly
obeys basic physical requirements, including detailed balance.
In particular relaxation is  described by introducing the 
density matrix  ($\sigma$) written on the basis of the eigenstates of
$H(0)$. On this basis the equilibrium density matrix is diagonal, with
diagonal elements describing the Boltzmann populations of the
relevant states. Relaxation dynamics of digonal elements of $\sigma$ 
are only affected by inelastic scattering and are written as:\cite{Mukamel,Boyd}
\begin{equation}
  (\dot\sigma_R)_{kk}=\sum_m \gamma_{km}\sigma_{mm} -\sum_m \gamma_{mk}\sigma_{kk}
\end{equation}
where $\gamma_{km}$ measures the probability of the transition from
$m$ to $k$ as due to inelastic scattering events. 
Off diagonal elements of the relaxation matrix may be written as:
\begin{equation}
  (\dot\sigma_R)_{km}=-\Gamma_{km}\sigma_{km}
\end{equation}
where 
\begin{equation}
  \Gamma_{km}=(\gamma_{kk}+\gamma_{mm})/2+\gamma'_{km}
\end{equation}
accounts for both inelastic scattering events, via the population
inverse lifetimes $\gamma_{kk}=\sum'_m\gamma_{mk}$,
 and for elastic scattering events via  
$\gamma'_{km}$ that describes the loss of coherence due to pure dephasing
phenomena.\cite{Mukamel,Boyd}

The total work  exchanged by the junction, 
$Tr(\dot\sigma_R \hat H)$, has 2 non-vanishing contributions:
$W=-\lambda Tr(\dot\sigma_R \hat{j})$
that represents the work done on the junction to sustain the current 
and a second contribution, $W_d=Tr(\dot \sigma_RH_0)$, that measures
the work dissipated by the junction.
Since the current operator, $\hat j$, has vanishing diagonal
elements on the real eigenstates of $H(0)$,
$W$ is only affected by the relaxation of off-diagonal elements of the density
matrix. As a result, both elastic and inelastic scattering
events affect $W$. This is a physically relevant result; in fact, both
phenomena concur to build up the resistance in the junction.
On the opposite,
the power dissipated by the junction, $W_d=Tr(\dot \sigma_RH_0)$,  is
only affected by inelastic scattering events. It represents just a
fraction of the total power dissipated on the junction, thus
suggesting that some dissipation must occur at the 
electrodes.\cite{Anna} }

{\color{blue} While the relaxation model is fairly crude and requires invoking
electrodes to actually balance the energy, it has the advantage of
setting on a firm basis the relation between the molecular resistance
and dissipation phenomena. In particular we notice that the proposed model
properly predicts the maximum conductivity of the simplest junction as
$e^2/\hbar$ as a result of the maximum lifetime of the electron within
the junction which cannot be longer than the time required to the
electron to cross the junction.\cite{Anna} This observation drives us
to the non-trivial role of electrodes on relaxation dynamics.}
In the case of very weak contacts, the presence of metallic surfaces
close to the molecular junction is expected to open new channel for
energy dissipation then reducing the population
lifetimes.\cite{lifetime}  On the
opposite, in the strong contact regime the mixing of the discrete
molecular eigenstates with the continuum of states of the metallic
leads is responsible for increased dephasing rates.\cite{Datta3}
 A very simple
model for the relaxation matrix can then be obtained in the strong
contact regime, where  depopulation contributions to $\Gamma_{km}$
can be neglected, and in the hypothesis that the effects of the
electrodes on the dephasing rates is similar on all states, so that $\Gamma_{km}=\Gamma$.
 In this hypothesis $W=\lambda \Gamma I$ and the potential drop is
 simply proportional to $\lambda$: $V=\lambda
\Gamma$.\cite{Anna} 
Without loss of generality we set $\Gamma=1$ in the following. 

\section{Results and Discussion}

 We consider two different choices for the energies of the
auxiliary sites: in the first model (model a) we set $\epsilon_1$ =
-$\epsilon_4$ to mimic groups with opposite electron donating/accepting characteristics.
In the second model (model b) we set $\epsilon_1$ =
$\epsilon_4$ mimicking two equivalent side-groups added to the main
chain (R=R' in Fig. 1(b)).  Quite interestingly, the second
model can also describe the effects of charge injection/depletion in  the
junction as a result of an applied gate voltage. 
The color maps in Fig. 2  show the differential conductance ($\frac
{\partial I}{\partial V}$) calculated  as a function of the potential drop ($V$) 
and of the energies of the auxiliary sites, $\epsilon_1 = -\epsilon_4$
(left panel) and $\epsilon_1 = \epsilon_4$ (right panel) for $U=5$.
NDC is not observed in either case.
This result, confirmed at different
$U$ values,  is  not surprising. In fact  NDC suggests a
largely non-linear behavior of the junction,  with a bistable  $V(I)$
characteristic. Such a behavior requires cooperativity and hence the
introduction of some self-consistent interaction in the model.\\
\begin{figure}
\centering
\includegraphics[scale=0.35,angle=0.0]{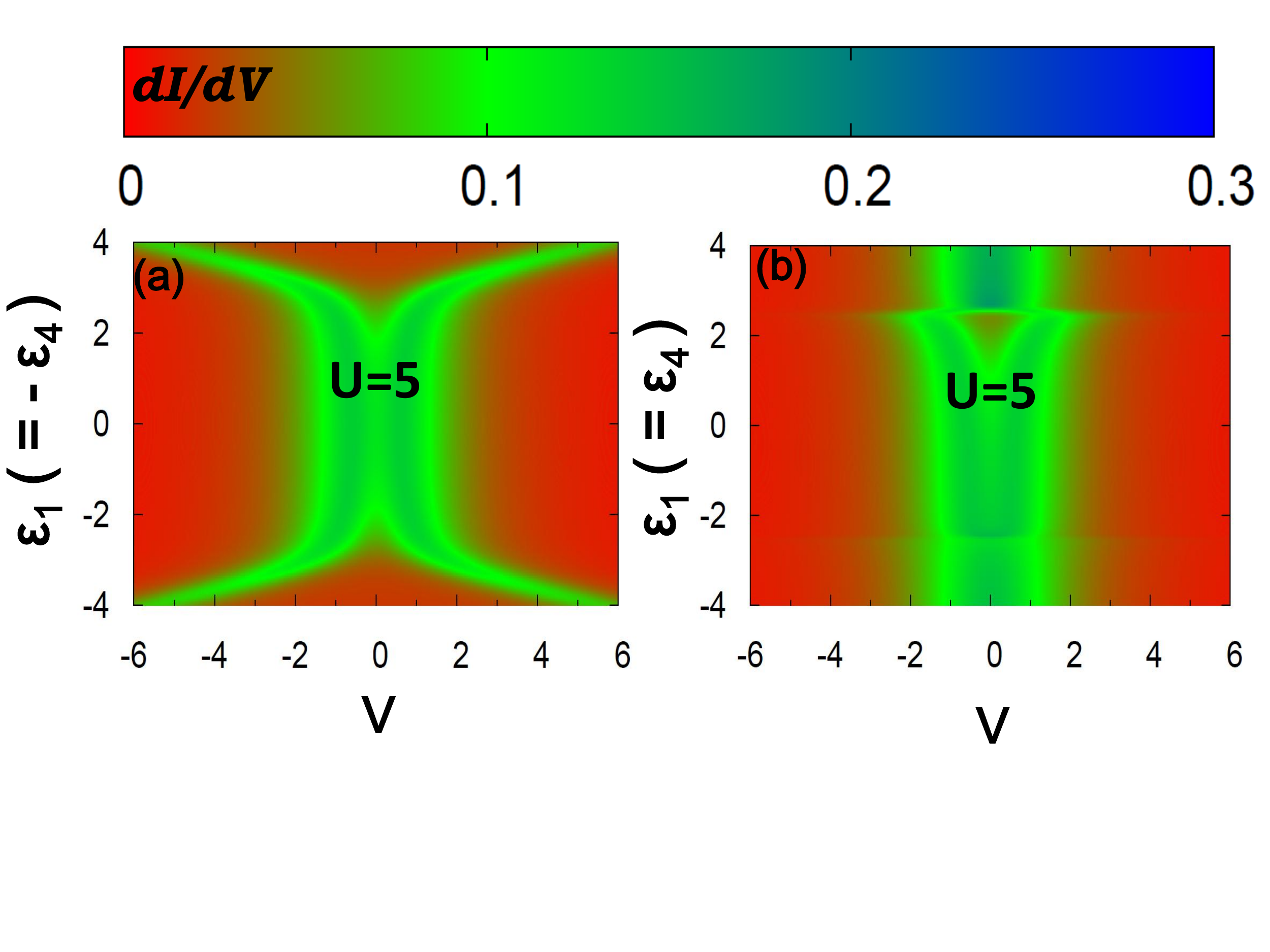}
\vspace*{-2.0cm}
\caption{\label{fig2} The color map shows differential conductance ($\frac
{\partial I}{\partial V}$) as a function of potential drop ($V$)
and of the energies of the auxiliary sites satisfying the relations (a) $\epsilon_1 =-\epsilon_4$ and
(b) $\epsilon_1 =\epsilon_4$. Results are shown for  $U$=5. }
\end{figure}

   With this requirement in mind, we modify the model to account for
possible effects of the voltage drop on the auxiliary sites. 
In fact, sites 1 and 4 do not play any active role in transport,  but
can be affected by the voltage drop at the junction. In particular,
being directly connected to sites 2 and 3 (device region), they can
experience at least a portion of the electrical field required to
sustain the current in the main junction. As a result, a term is added
to the Hamiltonian affecting the auxiliary-site energies as
follows:
\begin{equation}
x V (\hat{n_4}-\hat{n_1}), 
\label{correction}
\end{equation}
where  $x$ measures the fraction of the potential drop that is felt by
sites 1 and 4. We observe that  only the energy of auxiliary sites must
be explicitly  modified by the potential: the effect of the potential
on the junction (sites 2 and 3) is already accounted for by
the Hamiltonian in Eq. \ref{hlambda}.
 The  correction  term in Eq. \ref{correction} introduces cooperativity
in the model: the on-site energies, $\epsilon_1$ and $\epsilon_4$, do
depend in fact on the current flowing through the junction that, in
 turn is affected by the on-site energies. Color maps in Fig. 3,
obtained setting $x=0.5$ show NDC regions, demonstrating
the cooperative nature of the phenomenon. 
Fig. 3 summarizes the main results of this paper showing the
differential conductance calculated for  the two choices, (a)
$\epsilon_1=-\epsilon_4$ (left columns), and (b) $\epsilon_1=\epsilon_4$
(right columns) and $U=$5 and 2 (top and bottom panels,
respectively). NDC feature is definitely more
prominent in model b than in model a and  is suppressed
upon decreasing electron correlation strength, $U$ (this conclusion
is also supported by extensive calculations run at different $U$).
These findings  are in line with results obtained in the VC approach,
showing that strong on-site $e-e$ correlations favor NDC in double-quantum dot systems.~\cite{prakash}  
Note that, the ($\frac{\partial I}{\partial V}$) map (and the $I-V$
characteristics as well) are asymmetric because of 
the inherent asymmetry in the device region, with 
sites 2 and 3 having different on-site energies.
\begin{figure}
\centering
\includegraphics[scale=0.35,angle=0.0]{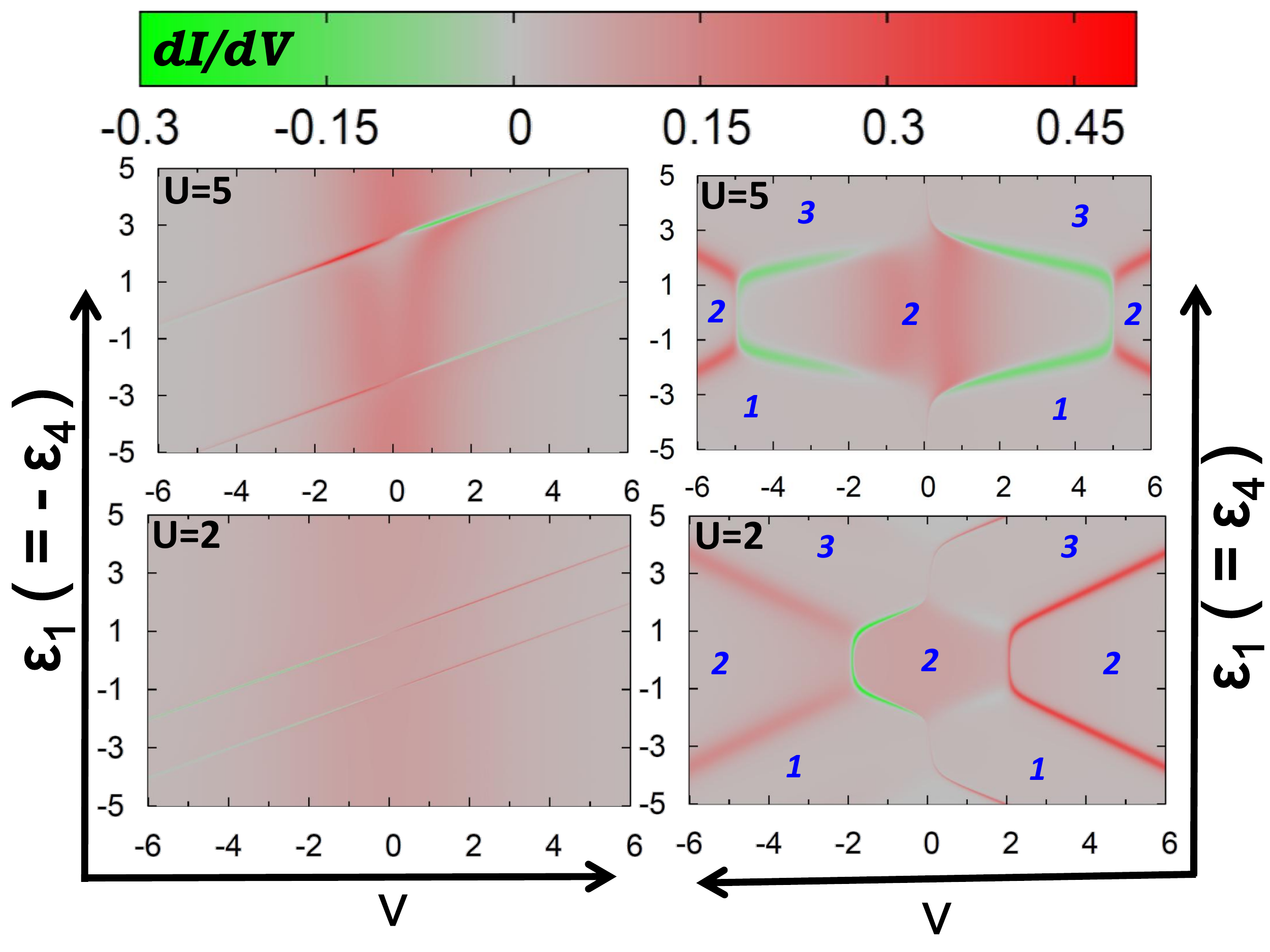}
\caption{\label{fig3} Color map showing differential conductance ($\frac
{\partial I}{\partial V}$) as a function of potential drop ($V$) 
and the energies of the auxiliary sites for  $\epsilon_1 =-\epsilon_4$
(left panels) and  $\epsilon_1 =\epsilon_4$ (right panels), for two
different $U$ values. The blue numbers in the right panels show the
approximate value of the number of electrons in the junction, $n$, in
the different regions. The same number is constant ($n=2$) in the left
panels. }
\end{figure}

Due to symmetry reasons, the  number of electrons in the junction
 is constant in model (a), $n=2$. In model (b), $n$ is instead
variable.  Since  the
hybridization energy, $t$, is small with respect to on-site energies,
$n(V,\epsilon_1)$  assumes almost integer and constant values
(shown as italic numbers in the figures) in different regions of the $V,
\epsilon_1$ plane connected by very sharp borders. Quite impressively,
NDC features are observed for  model (b) just at these borders, suggesting a strong
relation between NDC  and sharp variations of $n$. Upon
decreasing $U$, the variation of $n$  becomes smooth
and the NDC regions  become less prominent.
Minor NDC features  are also observed in
model (a) with constant $n$, in regions where 
 the occupancy of the two auxiliary sites abruptly  changes from $\langle
\hat n_1-\hat n_4\rangle$ =0 to $\pm 2$. 
\begin{figure}
\centering
\includegraphics[scale=0.4,angle=0.0]{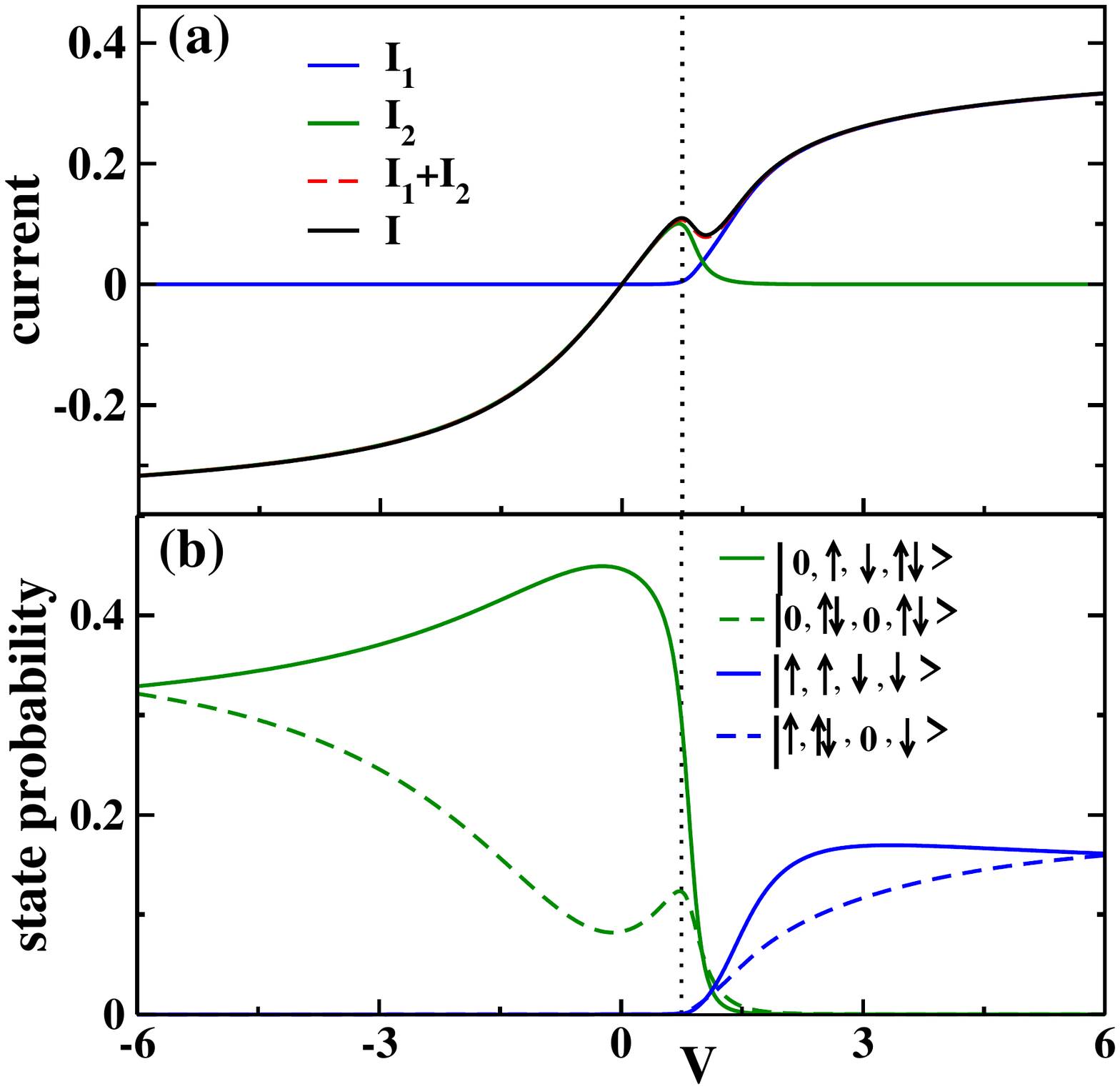}
\caption{\label{fig6} Results for $\epsilon_1=-\epsilon_4=3 $, and
  $U=5$. The black line in the top panel shows the characteristic
  curve. The blue and green  lines in the same panel show the
  two major contributions to the current  as discussed in the
  text, the red dashed line shown the sum of the two contribution, and
  it is almost indistinguishable from the total current. In the bottom panel
  blue (green) lines show the weight of the two states responsible for
  the $I_1$ ($I_2$), as discussed in the text.
 The black dotted line is shown as a  guide for the eyes}
\end{figure}

  To obtain a more detailed  picture of the  microscopic origin of NDC, we 
disentangle the contributions to the current from the different
states. {\color{blue} Specifically we work a real-space basis, where
  states are defined assigning electrons with a specified spin to 
each site in the junction. A generic state $|p\rangle=|0,\uparrow,\downarrow,
\downarrow\uparrow\rangle$  then graphically represents the 
state where no electrons are found in the first site, one electron
with spin $\alpha$ and one electron with spin $\beta$ are located on sites 2
and 3, respectively, and site 4 is doubly occupied. In more formal
language: $|0,\uparrow,\downarrow, \downarrow\uparrow\rangle= a^\dagger_{2,\alpha}
a^\dagger_{3,\beta}  a^\dagger_{4,\beta}a^\dagger_{4,\alpha}|0\rangle$
where $a^\dagger_{i,\sigma}$ are the electron creation operators
entering the Hamiltonian in eq. \ref{hami} and  $|0\rangle$ is vacuum state-
The current-carrying ground state has complex coefficients $c_p$ 
in the real space basis, $|G(\lambda)\rangle=\sum _p c_p
|p\rangle$  and the expectation value of the current
operator  can be written as $I=\sum_{pr}I_{pr}$, where
$I_{pr}=c_pc_r^*j_{pr}$, and $j_{pr}$ is an off-diagonal element of the current
operator on the chosen basis  $j_{pr}=\langle p|\hat{j}|r\rangle$ 
(diagonal elements of $\hat j$ vanish in the chosen basis).}
For the sake of clarity, we set $U=5$ in the 
following discussion. Fig 4 shows the results relevant to model (a) with
 $\epsilon_1=-\epsilon_4=3$. In this case, as noticed before,
 $n=2$. Two main contributions to $I$ can be recognized as 
$I_1=I_{|\uparrow,~\uparrow\downarrow,0,~ 
\downarrow>|\uparrow,~\downarrow,\uparrow,~ \downarrow>}+
I_{|\downarrow,~\uparrow\downarrow,0,~ 
\uparrow>|\downarrow,~\uparrow,\downarrow,~ \uparrow>}$
and
$I_2=I_{|0,~\uparrow\downarrow,~0,~\uparrow\downarrow>|0,~\uparrow,~\downarrow,~\uparrow\downarrow>}+
I_{|0,~\uparrow\downarrow,~0,~\uparrow\downarrow>|0,~\downarrow,~\uparrow,~\uparrow\downarrow>}$. Quite
interestingly,
 as shown in Fig. 4, the  sum of these two contributions quantitatively
matches with the 
total $I(V)$ characteristic.
The states involved in $I_1$ and in $I_2$ only differ by the
occupation  at 1st and 4th sites: the states involved in $I_1$ show
single  occupancy of  sites 1 and 4, while the states involved in
$I_2$ have zero occupation at site 1 and double occupation at site 4.
The position and height of the NDC peak of $I_2(V)$ exactly matches with
the NDC peak of the $I(V)$ characteristics:  the fall of $I_2$ and the rise of $I_1$ at $V \sim$ 0.75 
causes the NDC feature in $I(V)$ characteristics. 
For $V >$ 0.75, $I_2$ starts decreasing to zero, while $I_1$ increases 
to meet $I_2$ at $V$=1.04 resulting a valley in $I(V)$. 
So, for $V >$ 1.04, $I_1$ dominates over $I_2$ resulting in
further increase of total $I$.\\
\begin{figure}
\centering
\includegraphics[scale=0.4,angle=0.0]{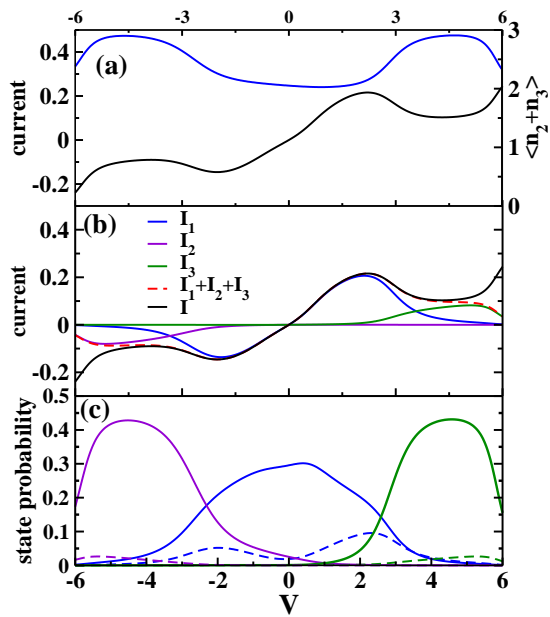}
\caption{\label{fig5ap} Results for $\epsilon_1=\epsilon_4=2 $, and
  $U=5$. Panel (a): Characteristic  curve (black line) and total
  number of electrons in the junction (blue line). Panel (b) the three
  main contributions to the current in the NDC regions,
as discussed in the text. The red dashed lines shows their sum to be
compared to the total current (black line).
Panel (c) shows the weight of states contributing to $I_1$ (blue
lines), $I_2$ (violet lines) and $I_3$ (green lines).
More specifically the blue continuous
and dashed line refer to states $|\uparrow , \downarrow, \uparrow ,
\downarrow \rangle $ and $|\uparrow , \downarrow \uparrow , 0,
\downarrow \rangle $, respectively;  the green continuous and dashed
lines refer to $|\downarrow , \downarrow \uparrow , \uparrow,
0 \rangle $, and $|\downarrow , \uparrow, \downarrow \uparrow,
0 \rangle $, respectively; the violet continuous and dashed lines
refer to $|0 , \downarrow \uparrow , \uparrow,\downarrow\rangle $, and
$|0 , \uparrow, \downarrow \uparrow,
\downarrow \rangle $, respectively. }
\end{figure}

 The bottom panel of Fig. 4 shows the
weight of  the states that mainly contribute to the
current, as shown in the figure label.
The probability of states contributing to $I_1$(blue lines) 
vanishes for $V<0.75$: $I_1$=0 in that region. $I_1$ instead
contributes significantly to total $I$ for  $V>0.75$, where the
contribution from $I_2$ becomes negligible.
The current flows only through the central 
bond, and, in a system with $n=2$ and $\epsilon_2=-\epsilon_3=2$, it 
is associated with off-diagonal elements of $\hat j$ mixing up states where 
both  electrons reside on site 2 or are equally distributed in sites 2 and 3. NDC is
observed when the occupancy of the auxiliary sites changes:
the NDC feature occurs when the population of states 
contributing to $I_2$, characterized by $\langle \hat n_4- \hat n_1\rangle =2$, 
start decreasing at the expense of the increasing population of states
contributing to $I_1$, characterized by $\langle \hat n_4- \hat n_1 \rangle=0$.\\ 

Model b is more complex, as shown in 
Fig. 5, collecting  the results obtained for $\epsilon_1=\epsilon_4=2$ and $U=5$.
In this case  a variable number of electrons is found in the junction, as
shown by the blue lines  in Fig. 5a. The 
$I-V$ characteristic (black line) shows NDC
features both in the negative and positive $V$ regions, with a more
prominent feature at  $V\sim 2$.
In the region of interest for NDC the current is mainly due to three
specific contribution, as shown in Fig. 5 b. The three contributions
are:
$I_1=I_{|\uparrow, ~\uparrow\downarrow, ~0,
  ~\downarrow\rangle|\uparrow, ~\downarrow,
  ~\uparrow, ~\downarrow\rangle  } 
+I_{|\downarrow, ~\uparrow\downarrow, ~0,
  ~\uparrow\rangle|\downarrow, ~\uparrow,
  ~\downarrow, ~\uparrow\rangle } $,
$I_2=I_{|0,~\uparrow\downarrow, ~\uparrow, \downarrow \rangle
|0, ~\uparrow, ~\uparrow\downarrow, ~\downarrow\rangle }+
I_{|0,~\uparrow\downarrow, ~\downarrow, \uparrow \rangle
|0, ~\downarrow, ~\uparrow\downarrow, ~\uparrow\rangle }$
and 
$I_3=I_{|\downarrow,~\uparrow\downarrow, ~\uparrow, 0 \rangle
|\downarrow, ~\uparrow, ~\uparrow\downarrow, ~0\rangle }+
I_{|\uparrow,~\uparrow\downarrow, ~\downarrow, 0 \rangle
|\uparrow, ~\downarrow, ~\uparrow\downarrow, ~0\rangle }$.
The states involved in $I_1$ have two electrons in
the junction, $n=2$, while  the
states contributing to $I_2$ and $I_3$ have $n=3$. 
As expected, $I_2$ and $I_3$ contributions become sizable for large
negative and positive $V$, respectively.
However it is clear from the figure
that the NDC features observed in the characteristic $I(V)$ are only
due to the  NDC features 
observed  in the $I_1(V)$ contribution to the total current. 
For a better understanding, Fig. 5c shows  the probability of states
playing a main role in transport. $I_1$ drops at the NDC features because the relevant states,
having two electrons in the junction, loose their weight and the
increasing contributions of either
$I_2$  in the negative $V$ region or of  $I_3$  
 in the positive $V$ region, related to states with
$n=3$, are not able to fully compensate for the loss.

\section{conclusions}

The current constrained approach offers an alternative picture of
transport in nanojunctions with respect to the widespread
voltage-constrained approaches. In the CC approach Lagrange
multipliers are used to force the system into a current carrying
state, then avoiding any explicit reference to semi-infinite
electronic reservoirs. The method then applies quite naturally to
describe transport in systems with strongly correlated
electrons. However, the total number of electrons taking part in
transport is fixed in CC approaches, and this has hindered so far  
the application of the
method to problems where  charge injection/depletion in the junction is
important. {\color{blue}This problem could be solved introducing an effective
relaxation model that, accounting for the exchange of electrons
between the junction and the electrodes, allows the system to relax
towards states with different number of electrons. This approach
however requires a fairly complex model for relaxation dynamics. Here
we borrow from the field of optical spectroscopy  a simple
phenomenological  model for relaxation that maintains constant the
number of electrons, and
overcome the limitation of fixed number of electrons}
connecting one or more auxiliary sites to the
main junction: while the current does not flow through the auxiliary sites,
they exchange electrons with the junction,  allowing for a
variable  number of electrons taking part in transport.

Along these lines we attacked the problem of NDC in a very simple
model with a  two-site junction connected to two auxiliary sites.
NDC is a highly non linear phenomenon: the bistable behavior of the
$V(I)$ curve at NDC can
only be observed in largely cooperative systems.
We introduced cooperativity imposing a dependence of
the on-site energies of the auxiliary sites on the voltage drop:
since  the voltage
required to support the current depends in turn  on the energy of the
auxiliary sites, the problem becomes self-consistent and the model supports NDC.
 Quite interestingly, dangling bonds (side groups) connected to the main
current-carrying unit have recently been discussed as responsible for
non-linearity in the characteristic curves. Particularly side groups
may be responsible for quantum interference phenomena that drastically
suppress current.\cite{ultimi} Our study  suggests that  a proper choice of
the chemical nature of the side groups can not only reduce the
current, but eventually  lead to NDC.

Here we just  investigated an asymmetric junction with weak bonds
and  a
grand-total of four electrons, in two different cases. In the first case we
choose equal and opposite on-site energies for the two auxiliary sites
leading to a constant number of electrons in the junction
($n=2$). Minor NDC features are observed in this case, mainly related
to an abrupt variation of the occupation of the auxiliary sites. More
prominent NDC features are observed in the second model, where, having fixed the
energies of the auxiliary sites to the same value, the number of
electrons involved in the transport is variable. NDC is 
related in this case to an abrupt variation with the applied potential
of the number of electrons in the
junction. In all cases,  large $U$ and small $t$ are needed to observe 
NDC: weak bonds and large correlations in fact ensure the abrupt
variation in the nature of the current carrying state
$|G(\lambda)\rangle$, as required for to observe NDC.

\section{Acknowledgments}
PP acknowledges the CSIR for a research fellowship, SKP acknowledges research support 
from CSIR and DST, Government of India and AP acknowledges support
from the Italian Ministry of Foreign Affairs  and Parma University.



\begin{thebibliography}{50}
\bibitem{Ratner} A. Nitzan and M. A. Ratner, Science {\bf 300}, 1384 (2003);
W. A. Svec, M. A. Ratner, and M. R. Wasielewsk, Nature {\bf 396}, 60 (1998);
R. G. Endres, D. L. Cox, and R. R. P. Singh, Rev. Mod. Phys. {\bf 76}, 195 (2004);
S. S. Mallajosyula, P. Parida, and S. K. Pati, J. Mat. Chem. {\bf 19}, 1761 (2009).
\bibitem{Sanvito} T. Rueckes, K. Kim, E. Joselevich, A. R. Rocha, V. M. Garcasurez,
S. W. Bailey, C. J. Lambert, J. Ferrer, and S. Sanvito, Nature Materials {\bf 4}, 335 (2005);
A. Zutic, J. Fabian, and S. D. Sarma, Rev. Mod. Phys., {\bf 76}, 323 (2004);
P. Parida, A. Kundu, and S. K. Pati, Phys. Chem. Chem. Phys. DOI: 10.1039/c004653c (2010);  
S. K. Pati, J. Chem. Phys. {\bf 118}, 6529 (2003). 
\bibitem{Esaki}L. Esaki, Phys. Rev. {\bf 109}, 603 (1958). 
\bibitem{Tour} J. Chen, M. A. Reed, A. M. Rawlett, and J. M. Tour, Science {\bf 286}, 1550 (2001);
J. Chen, W. Wang,  M. A. Reed, A. M. Rawlett, D. W. Price, and J. M. Tour, Appl. Phys. Lett. {\bf 77}, 1224 (2000).
\bibitem{Don} Z. J. Donhauser, B. A. Mantooth, K. F. Kelly, L. A. Bumm, J. D. Monnell, J. J. Stapleton, D. W. Price, Jr,
A. M Rawlett, D. L. Allara, J. M. Tour, and P. S. Weiss, Science {\bf 292}, 2303 (2001).
\bibitem{Seminario} J. M. Seminario, A. G. Zacarias, and J. M. Tour, J. Am. Chem. Soc. {\bf 122}, 3015 (2000);
J. M. Seminario, A. G. Zacarias, and P. A. Derosaet, J. Chem. Phys. {\bf 116}, 1671 (2002).
\bibitem{Cornil} J. Cornil, Y. Karzazi, and J. L. Bredas, J. Am. Chem. Soc. {\bf 124}, 3516 (2002);
S. Lakshmi and S. K. Pati, J. Chem. Phys. {\bf 121}, 11998 (2004).
\bibitem{rpati} R. Pati and S. P. Karna, Phys. Rev. B {\bf 69}, 155419 (2004).
\bibitem{lakshmi_PRB} S. Lakshmi and S. K. Pati, Phys. Rev. B {\bf 72}, 193410 (2005).
\bibitem{lakshmi_jpc} S. Lakshmi, S. Dutta, and S. K. Pati, J. Phys. Chem. C {\bf 112}, 14718 (2008).
\bibitem{Cuniberti1}  R. Gutierrez, G. Fagas, G. Cuniberti, F. Grossmann, R. Schmidt, and K. Richter,
Phys. Rev. B {\bf 65}, 113410 (2002).
\bibitem{ranjit1} R. Pati, M. McClain, and A. Bandyopadhyay, Phys. Rev. Lett. {\bf 100}, 246801 (2008). 
\bibitem{Ono} K. Ono, D. G. Austing, Y. Tokura, and S. Tarucha, Science {\bf 297}, 1313 (2002).
\bibitem{tarucha} K. Ono and S. Tarucha, Phys. Rev. Lett. {\bf 92}, 256803 (2004).
\bibitem{Thielmann} A. Thielmann, M. H. Hettler, J. Konig, and G. Schon, Phys. Rev. B {\bf 71}, 045341 (2005).
\bibitem{Hettler1} M. H. Hettler, W. Wenzel, M. R. Wegewijs, and H. Schoeller, Phys. Rev. Lett. {\bf 90}, 076805 (2003).
\bibitem{wegewis} M. R. Wegewis, M. H. Hettler, W. Wenzel, and H. Schoeller, Physica E, {\bf 18}, 241 (2003).
\bibitem{Hung}  N. V. Hung, N. V. Lien, and P. Dollfus, Appl. Phys. Lett. {\bf 87}, 123107 (2005).
\bibitem{Cuniberti} B. Song, D. A. Ryndyk, and G. Cuniberti, Phys. Rev. B {\bf 76}, 045408 (2007).
\bibitem{Aghassi} J. Aghassi, A. Thielmann, M. H. Hettler, and G. Sch\"on, Phys. Rev. B
{\bf 73}, 195323 (2006).
\bibitem{Paulsson} M. Paulsson and S. Stafstrom, Phys. Rev. B {\bf 64}, 035416 (2001).
\bibitem{Datta2} B. Muralidharan and S. Datta, Phys. Rev. B {\bf 76}, 035432 (2007).
\bibitem{Bhaskaran} B. Muralidharan, A. W. Ghosh, and S. Datta, Phys. Rev. B {\bf 73}, 155410 (2006).
\bibitem{Datta3} S. Datta, Nanotechnology {\bf 15}, S433 (2004).
\bibitem{IEEE} B. Muralidharan, A. W. Ghosh, S. K. Pati, and S. Datta, IEEE
Transac. on Nanotech. {\bf 6}, 536 (2007).
\bibitem{prakash} P. Parida, S. Lakshmi, and S. K Pati, J. Phys.: Cond. Matt. {\bf 21}, 095301 (2009).
\bibitem {Kohn1}  A. Kamenev and W. Kohn, Phys. Rev. B {\bf 63}, 155304 (2001).
\bibitem{Mera}   P. Bokes,  H. Mera, and R. W. Godby, Phys. Rev. B {\bf
  72}, 165425 (2005).
\bibitem{Green}  F. Green, J. S. Thakur,and M. P. Das, Phys. Rev. Lett {\bf 92}, 156804 (2004);
M. P. Das and F. Green, J. Phys. Cond. Matt. {\bf 15}, L687 (2003).
\bibitem {Ng} T. K. Ng, Phys. Rev. Lett. {\bf 68}, 1018 (1992).
\bibitem{Magnus} W. Magnus and W. Schoenmaker, Phys. Rev. B {\bf 61}, 10883 (2000).
\bibitem{Kosov1} D. S. Kosov, J. Chem. Phys. {\bf 116}, 6368 (2002).
\bibitem {Kosov2}D. S. Kosov, J. Chem. Phys. {\bf 120}, 7165 (2004).
\bibitem {Burke} K. Burke, R. Car, and R. Gebauer, Phys. Rev. Lett. {\bf 94}, 146803 (2005).
\bibitem{Anna}   A. Painelli, Phys. Rev. B {\bf 74}, 155305 (2006).
\bibitem{Kohn2}  W. Kohn, Phys. Rev. {\bf 133}, A171 (1964).
\bibitem{maldague} {\color{red}P. F. Maldague, Phys. Rev. B {bf 16}, 2437 (1977).}
\bibitem{mukamelRMP} {\color{red} M. Esposito, U. Harbola, S. Mukamel,
  Rev. Mod. Phys. {\bf 81}, 1665 (2009). }
\bibitem{Mukamel} S. Mukamel, Principles of Nonlinear Optical Spectroscopy (Oxford
University Press, New York, 1995).
\bibitem{Boyd} R. W. Boyd, Nonlinear Optics (Academic Press, New York, 2003).
\bibitem{lifetime} {\color{red}R. R. Chance, A. Prock, and R. Silbey,
  Adv. Chem. Phys. {\bf 37}, 1     (1978);  M. Galperin and A. Nitzan, 
Phys. Rev. Lett. {\bf 95}, 206802 (2005).}
\bibitem{ultimi} {\color{red}G.C. Solomon, D. Q. Andrews, R. H. Goldsmith, T. Hansen,
     M. R. Wasielewski, R. P. Van Duyne, M. A. Ratner,
     J. Am. Chem. Soc. {\bf  130}, 17301 (2008); 
 G. C. Solomon, D.Q. Andrews, R. P. Van Duyne,
  M. A. Ratner, ChemPhysChem {\bf  10}, 257 (2009).}
\end{thebibliography}
\end{document}